\documentclass[11pt]{article}
\usepackage{amsmath,epsf,amssymb,latexsym,enumerate,cite,shadow,array,color}
\usepackage{slashed}
\usepackage{graphicx,wasysym}
 \pdfoutput=1

\DeclareFontFamily{U}{rsf}{} \DeclareFontShape{U}{rsf}{m}{n}{
  <5> <6> rsfs5 <7> <8> <9> rsfs7 <10-> rsfs10}{}
\DeclareMathAlphabet\Scr{U}{rsf}{m}{n} \makeatletter
\@addtoreset{equation}{section} \makeatother

\def\be{\begin{equation}}
\def\ee{\end{equation}}
\def\ba{\begin{array}}
\def\ea{\end{array}}

\newcommand{\bea}{\begin{eqnarray}}
\newcommand{\eea}{\end{eqnarray}}

\def\K{K{\"a}hler}
\newcommand{\ft}[2]{{\textstyle\frac{#1}{#2}}}

\def\rme{{\rm e}}
\usepackage{ifpdf}
\ifx\pdfoutput\undefined
   \pdffalse
\else
   \pdfoutput=1
   \pdftrue
  \usepackage[pdftex]{hyperref}
\newcommand{\rf}[1]{(\ref{#1})}

\begin{document}

\begin{titlepage}

\

\

\begin{center}
{\LARGE \textbf{Universality Class in Conformal Inflation
\vskip 0.8cm }}

\

{\bf Renata Kallosh} and {\bf Andrei Linde}

{\sl Department of Physics and SITP, Stanford University, Stanford, California
94305 USA}\\
\end{center}
\vskip 1.5 cm

\

\begin{abstract}

We develop a new class of chaotic inflation models with spontaneously broken conformal invariance.  Observational consequences of a broad class of such models are stable with respect to strong deformations of the scalar potential.  This universality is a critical phenomenon near the point of enhanced symmetry, $SO(1,1)$, in case of conformal inflation. It appears because of the exponential stretching of the moduli space and the resulting exponential flattening of scalar potentials upon switching from the Jordan frame to the Einstein frame in this class of models. This result resembles stretching and flattening of inhomogeneities during inflationary expansion. It has a simple interpretation in terms of velocity versus rapidity near the \K\ cone in the moduli space, similar to the light cone of special theory of relativity.  This effect makes inflation possible even in the models with very steep potentials.   We describe conformal and superconformal versions of this cosmological attractor mechanism.

 \end{abstract}

\vspace{24pt}
\end{titlepage}

%{\parskip -5pt \tableofcontents}

\tableofcontents

\newpage

\section{Introduction}

This paper extends the series of our recent papers \cite{Kallosh:2013pby,Kallosh:2013lkr} where we developed a superconformal approach to the description of a broad class of models favored by observational data from WMAP9 \cite{Hinshaw:2012aka} and Planck2013 \cite{Ade:2013rta}. The superconformal approach to cosmology is based on  earlier papers \cite{Kallosh:2000ve,Ferrara:2010yw,Ferrara:2010in}.

Recent observational data  attracted attention of cosmologists to a class of inflationary models which made very similar observational predictions, even though their formulations could  seem entirely unrelated to each other. This includes the early model \cite{Starobinsky:1980te}, with main observational predictions derived in \cite{Mukhanov:1981xt}. The same observational predictions were made in the context of chaotic inflation in the theory $\lambda\phi^{4}$ \cite{Linde:1983gd} with non-minimal coupling to gravity ${\xi\over 2}\phi^{2} R$ with $\xi > 0$ \cite{Salopek:1988qh,Sha-1},  in its supersymmetric extensions in \cite{Einhorn:2009bh,Ferrara:2010yw,Lee:2010hj,Ferrara:2010in}
and in a certain limit of the model with the Higgs potential with $\xi < 0$ \cite{Linde:2011nh}. Recently another set of models with similar predictions was identified in \cite{Ellis:2013xoa,Kallosh:2013pby,Kallosh:2013lkr,Buchmuller:2013zfa}. 

Relation of these models to each other remained quite mysterious. In our recent set of papers \cite{Kallosh:2013pby,Kallosh:2013lkr} we proposed that spontaneously broken superconformal symmetry is useful to describe these and other inflationary models favored by observations. 
In this paper we will describe a new class of models of inflation suggested by the framework of conformal and superconformal symmetry. These new models  form a universality class. In terms of the observational data, they all have an attractor point
\be
1 -n_{s} =2/N\, , \qquad r = 12/N^{2} 
\label{attractor}\ee
in the leading approximation in $1/N$, where $N$ is the number of e-folding of inflation. This class of models is very broad; it includes the Starobinsky  model as well as the Higgs model with $\xi <0$. The same attractor point  \rf{attractor} corresponds to Higgs inflation type models with $\xi> 0$, but they form a different universality class, as discussed in \cite{Kallosh:2013pby}.

 The class of models we study depends on two scalar fields. The action  has a local conformal symmetry, one field $\chi$ we call a conformon, following \cite{Kallosh:2000ve}, where the superconformal approach to cosmology was first developed.  The other field $\phi$ is an inflaton. Local conformal symmetry for these two fields in absence of a potential has an unbroken global $SO(1,1)$ symmetry. The potential can be added which preserves $SO(1,1)$ symmetry as well as a local conformal symmetry. This form a critical point of the system, which is a dS/AdS space, depending on the sign of the potential.
 
 A small deformation of $SO(1,1)$ symmetry for all of our models corresponds to deforming the $SO(1,1)$ conformally invariant potential by an arbitrary function $F(\phi/\chi)$, which is a unique way of preserving local conformal symmetry.  The simplest representative of this class of theories is $F(\phi/\chi) \sim (\phi/\chi)^{2}$; we called it the T-Model for the reasons to become apparent later. As we have shown in \cite{Kallosh:2013lkr}, a more complicated function $F(\phi/\chi) \sim {(\phi/\chi)^{2} \over 1 +(\phi/\chi)^{2}}$ corresponds to the Starobinsky model. We will show that these models and their numerous generalizations belong to the same universality class of slightly $SO(1,1)$ deformed models, which has an attractor point \rf{attractor}.

 In section 2 of this paper  we present examples of  actions with local conformal symmetry in presence of the conformon field,  which lead to dS or AdS spaces when conformal symmetry is spontaneously broken. We use two types of gauges to fix conformal symmetry, which lead to the same results in the Einstein frame.  These gauges are related to each other as velocity and rapidity in special theory of relativity and have different advantages for various purposes, which we use later in more interesting models. In section 3 we introduce a new class of inflationary models, the basic one we call T-Model, as a  deviation from the $SO(1,1)$ invariant conformal models with de Sitter geometry.
 We find that  their cosmological evolution is universal in the first approximation in $1/N$ where $N$ is the number of e-foldings, due to an exponential stretching and flattening of the inflaton potential near the boundary of the moduli space. In section 4 we investigate more general class of conformal inflation models, which have an analogous attractor behavior for slow roll inflation parameters and flattening of the inflationary potentials.\footnote{
Other ways of flattening of the inflaton potentials were studied in \cite{Dong:2010in}.}
 In section 5 we describe a supersymmetric generalization of the universality class of conformal inflation, and in section 6 we present some comments on general emergence of critical phenomena in the superconformal approach to cosmology.

\section{de Sitter from  spontaneously broken conformal symmetry}\label{dSsection}

\subsection{The simplest conformally invariant one-field  model of dS/AdS space}

As a first step towards the development of the new class of inflationary models based on spontaneously broken conformal symmetry, consider a simple  conformally invariant model of gravity 
of a scalar field $\chi$ with the following Lagrangian
\begin{equation}
\mathcal{L} = \sqrt{-{g}}\left[{1\over 2}\partial_{\mu}\chi \partial_{\nu}\chi \, g^{\mu\nu}  +{ \chi^2\over 12}  R({g}) -{\lambda\over 4}\chi^4\right]\,.
\label{toy2aa}
\end{equation}
This
theory is locally conformal invariant under the following
transformations: 
\be \tilde g_{\mu\nu} = \rme^{-2\sigma(x)} g_{\mu\nu}\,
,\qquad \tilde\chi =  \rme^{\sigma(x)} \chi\ . \label{conf2aaa}
\ee
The field $\chi(x)$ is referred to as  a conformal compensator, which we will call `conformon.'   It has negative sigh kinetic term, but this is not a problem because it
can be removed from the theory by fixing the gauge symmetry
(\ref{conf}), for example by taking a gauge $\chi =\sqrt 6$. This gauge fixing can be interpreted as a spontaneous breaking of conformal invariance due to existence of a classical field $\chi =\sqrt 6$.

After fixing $\chi = \sqrt 6$ the interaction term $-{\lambda\over 4}\chi^4$  becomes a cosmological constant $\Lambda = {9}\lambda$
\be
\mathcal{L}=  \sqrt{-{g}}\,\left[{  R({g})\over 2}-{9\lambda}\right]\, .
\label{toy2b}
\ee
For $\lambda > 0$, this theory has a simple de Sitter solution 
\be\label{dssol}
a(t) = e^{Ht}  = e^{\sqrt{\Lambda/3}\ t}  = e^{\sqrt{3\lambda}\ t}.
\ee
Meanwhile for $\lambda < 0$ it becomes the AdS universe with a negative cosmological constant.

\subsection{The simplest conformally invariant two-field  model of dS/AdS space}

Consider now the model of two real scalar fields, $\phi$ and $\chi$, which has also an $SO(1,1)$ symmetry:
\begin{equation}
\mathcal{L}_{\rm toy} = \sqrt{-{g}}\left[{1\over 2}\partial_{\mu}\chi \partial^{\mu}\chi  +{ \chi^2\over 12}  R({g})- {1\over 2}\partial_{\mu} \phi\partial^{\mu} \phi   -{\phi^2\over 12}  R({g}) -{\lambda\over 4}(\phi^2-\chi^2)^{2}\right]\,.
\label{toy}
\end{equation}
This
theory is locally conformal invariant under the following
transformations: 
\be \tilde g_{\mu\nu} = \rme^{-2\sigma(x)} g_{\mu\nu}\,
,\qquad \tilde \chi =  \rme^{\sigma(x)} \chi\, ,\qquad \tilde \phi =  \rme^{\sigma(x)}
\phi\ . \label{conf}\ee 
The global $SO(1,1)$ symmetry is a boost between these two fields.

\subsection{  $SO(1,1)$ invariant conformal gauge: rapidity gauge $\chi^2-\phi^2=6$}
Because of this symmetry,   it is very convenient to choose an  $SO(1,1)$ invariant conformal gauge,
\be
\chi^2-\phi^2=6 \ ,
\ee
which reflects the $SO(1,1)$ invariance of our model. This gauge condition (we will refer to it as a `rapidity' gauge) represents a hyperbola which can be parametrized by a canonically normalized field $\varphi$ 
\be
\chi=\sqrt 6 \cosh  {\varphi\over \sqrt 6}\, , \qquad \phi= \sqrt 6 \sinh {\varphi\over \sqrt 6} \ .
\ee
This allows us to modify the kinetic term into a canonical one for the independent field $\varphi$
\begin{equation}
{1\over 2}\partial_{\mu}\chi \partial^{\mu}\chi  - {1\over 2}\partial_{\mu} \phi\partial^{\mu} \phi   \,  \quad \Rightarrow \quad {1\over 2}\partial_{\mu}\varphi \partial^{\mu}\varphi
\label{kin}
\end{equation}
and the non-minimal curvature coupling into a minimal one
\begin{equation}
{ \chi^2\over 12}  R({g})   -{\phi^2\over 12}  R({g})   \,  \quad \Rightarrow \quad { 1\over 2}  R({g})
\,.
\label{R}
\end{equation}

Using the standard terminology of special relativity, including the concept of rapidity,\footnote{Using the inverse hyperbolic function $\tanh \theta$, the rapidity $\theta$ corresponding to $\tanh^{-1} {v\over c}$. For low speed, $\theta$ is approximately ${v\over c}$. The speed of light c being finite, any velocity  is constrained to the interval $- c < v < c$ and the ratio ${v\over c}$ satisfies $-1<{v\over c}<1$. Since the inverse hyperbolic tangent has the unit interval $(-1, 1)$ for its domain and the whole real line for its range, the interval $- c < v < c$ becomes mapped onto $- \infty <\theta <\infty$. Thus rapidity is defined as the hyperbolic angle that differentiates two frames of reference in relative motion.}  we may define
$
 {\chi\over \sqrt 6}=\cosh {\varphi\over \sqrt 6} =\gamma \, ,    {\phi\over \sqrt 6}= \sinh {\varphi\over \sqrt 6}=\gamma {v\over c}
$
and
\be
\tanh {\varphi\over \sqrt 6}= {v\over c} \ .
\ee
For low speeds, rapidity and speed are proportional, but for high speeds, rapidity takes a larger value. 
We  see  that  the canonically normalized inflaton is related to `rapidity' $\theta=\tanh^{-1} {v\over c}$ as follows
\be
\theta = {\varphi\over \sqrt 6} \ .
\ee

In this gauge the Higgs-type potential ${\lambda\over 4}(\phi^2-\chi^2)^{2}$ turns out to be a cosmological constant $9\lambda$,
and our action \rf{toy} becomes
\begin{equation}\label{LE}
L = \sqrt{-g} \left[  \frac{1}{2}R - \frac{1}{2}\partial_\mu \varphi \partial^{{\mu}} \varphi -   9 \lambda  \right].
\end{equation}

This theory has the constant potential $V = 9\lambda$. Therefore this model can describe de Sitter expansion with the Hubble constant $H^{2} = 3\lambda$, for positive $\lambda$. For negative $\lambda$ in \rf{toy} the model can describe an AdS space. But unlike the one-field model, this model also describes a massless scalar field $\varphi$, which may have its kinetic and gradient energy. If one takes this energy into account, the universe approaches dS regime as soon as expansion of the universe makes the kinetic and gradient energy of the field $\varphi$ sufficiently small.

\subsection{$\chi(x) = \sqrt{6}$ conformal gauge}

One can also describe the model (\ref{toy}) differently. One may chose to fix the local conformal symmetry by using  the gauge $\chi(x) = \sqrt{6}$.The full
Lagrangian in the Jordan frame becomes
\begin{equation}
\mathcal{L}_{\rm total }= \sqrt{-{g_{J}}}\,\left[{  R({g_{J}})\over 2}\left(1-{ \phi^2\over 6}\right)-  {1\over 2}\partial_{\mu} \phi \partial^{\mu} \phi    - {\lambda\over 4} \left(\phi^2-6 \right)^{2}\right]\,.
\label{toy2}
\end{equation}
Now one can represent the same theory in the Einstein frame, by changing the metric $g_J$ and $\phi$ to a conformally related metric $g_{E}^{\mu\nu} = (1- \phi^{2}/6)^{{-1}}  g_{J}^{\mu\nu}$ and a new canonically normalized  field $\varphi$ related to the field $\phi$ as follows:
\be\label{field}
 \frac{d\varphi}{d\phi} = {1\over 1- {\phi^2/ 6}} \ .
\ee
 In new variables, the Lagrangian, up to a total derivative, is given by
\begin{equation}\label{LE1}
L = \frac{1}{2} \sqrt{-g} \left[ R - \frac{1}{2}g^{\mu\nu} \partial_\mu \varphi \partial_\nu \varphi -  V(\phi(\varphi)) \right].
\end{equation}
where the potential in the Einstein frame is 
\begin{equation}\label{eframe}
V(\phi) = { \lambda\over 4}  {\left(\phi^2-6 \right)^{2}\over \left(1-{ \phi^2\over 6}\right)^{2}} = 9\lambda \ .
\end{equation}
This finally brings the theory to the form which was found earlier in (\ref{LE}).

Here the relation between the gauges  $ \chi^2(x) -\phi^2(x)=6$ and $\chi(x) = \sqrt{6}$  corresponds to a relation between rapidity and velocity. The first in \rf{LE} brings us to an Einstein frame, the second one first brings us to a Jordan frame \rf{toy2}.  After switching from the Jordan frame to the Einstein one  finds a complete agreement. This is expected since rapidity is an alternative to speed description of a motion.

\section{Chaotic inflation from conformal theory: T-Model}

Now we will consider a  conformally invariant class of models
\begin{equation}
\mathcal{L} = \sqrt{-{g}}\left[{1\over 2}\partial_{\mu}\chi \partial^{\mu}\chi  +{ \chi^2\over 12}  R({g})- {1\over 2}\partial_{\mu} \phi\partial^{\mu} \phi   -{\phi^2\over 12}  R({g}) -{1\over 36} F\left({\phi/\chi}\right)(\phi^{2}-\chi^{2})^{2}\right]\,.
\label{chaotic}
\end{equation}
where $F$ is an arbitrary function of the ratio ${\phi\over \chi}$. When this function is present, it breaks the $SO(1,1)$ symmetry of the de Sitter model \rf{toy}. Note that this is the only possibility to keep local conformal symmetry and to deform the $SO(1,1)$ symmetry:  an arbitrary function $F\left({\phi/\chi}\right)$ has to deviate from the critical value $F\left({\phi/\chi}\right)=\rm const$,  where $SO(1,1)$ symmetry is restored.

This theory has the same conformal invariance as the theories considered earlier. As before, we may use the gauge $\chi^2-\phi^2=6$ and resolve this constraint in terms of the fields 
$\chi=\sqrt 6 \cosh  {\varphi\over \sqrt 6}$,  $ \phi= \sqrt 6 \sinh {\varphi\over \sqrt 6} $ 
and the  canonically normalized field $\varphi = {\phi\over\chi} = \tanh{\varphi\over \sqrt 6}$.
Our action \rf{chaotic} becomes
\begin{equation}\label{chaotmodel}
L = \sqrt{-g} \left[  \frac{1}{2}R - \frac{1}{2}\partial_\mu \varphi \partial^{\mu} \varphi -   F(\tanh{\varphi\over \sqrt 6}) \right].
\end{equation}
Note that asymptotically $\tanh\varphi\rightarrow \pm 1$ and therefore $F(\tanh{\varphi\over \sqrt 6})  \rightarrow \rm const$, the system  in large $\varphi$ limit evolves asymptotically  towards its critical point  where the $SO(1,1)$ symmetry is restored.

Since $F\left({\phi/\chi}\right)$ is an {\it arbitrary} function, by a proper choice of this function one can reproduce an {\it arbitrary} chaotic inflation potential $V(\varphi)$ in terms of a conformal theory with spontaneously broken conformal invariance. But this would look rather artificial. For example, to find a theory which has potential $m^{2}\varphi^{2}/2$ in terms of the canonically normalized field $\varphi$, one would need to use $F(\phi) \sim (\tanh^{{-1}} \phi)^{2}$, which is a possible but rather peculiar choice in the context of this class of theories.

Alternatively, one may think about the function $F\left({\phi/\chi}\right)$ as describing a deviation of inflationary theory from the pure cosmological constant potential which emerges in the theory (\ref{toy}) with the coupling $(\phi^{2}-\chi^{2})^{2}$. Therefore it is interesting to study what will happen if one takes the simplest set of functions $F\left({\phi/\chi}\right) = \lambda \left({\phi/\chi}\right)^{2n}$ as we did in the standard approach to chaotic inflation.

In this case one finds 
\begin{equation}\label{TModel}
V(\varphi) = \lambda_n\ {\tanh}^{2n}(\varphi/\sqrt6) .
\end{equation}
This is a basic representative of the universality class of models depending on $\tanh(\varphi/\sqrt6) $.  We will call it a T-Model, because it originates from different powers of $\tanh(\varphi/\sqrt6) $, and also because its basic representative $ \lambda_1\ {\tanh}^{2}(\varphi/\sqrt6)$ is the simplest version of the class of conformal chaotic inflation models considered in this paper. Moreover, as we will see, observational predictions of a very broad class of models of this type, including their significantly modified and deformed cousins, have nearly identical observational consequences, thus belonging to the same universality class. As Henry Ford famously exclaimed with respect to the T-Model Ford: ``Any customer can have a car painted any color that he wants so long as it is black.''

\begin{figure}[ht!]
\centering
\includegraphics[scale=1.2]{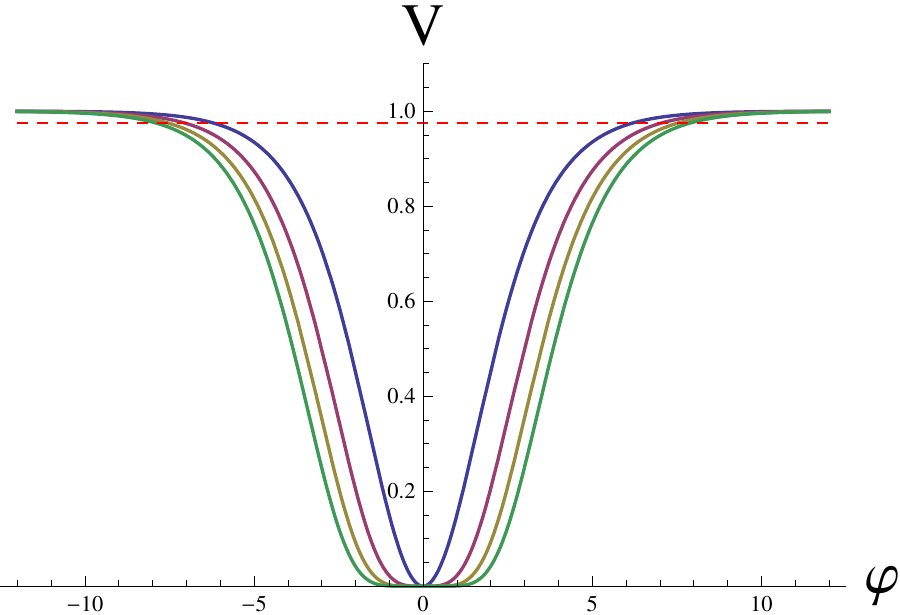}
\caption{Potentials for the T-Model inflation ${\tanh}^{2n}(\varphi/\sqrt6)$ for $n = 1,2,3,4$ (blue, red, brown and green,  corresponding to increasingly wider potentials).  We took $\lambda_{n} = 1$ for each of the potential for convenience of comparison. As we see, these potentials differ from each other quite considerably, especially at $\varphi \lesssim 1$: at small $\phi$ they behave as $\varphi^{2n}$. Nevertheless all of these models predict the same values $n_{s} =1-2/N$, $r = 12/N^{2}$, in the leading approximation in $1/N$, where $N\sim 60$ is the number of e-foldings. The points where each of these potentials cross the red dashed line $V = 1-3/2N = 0.975$ correspond to the points where the perturbations are produced in these models on scale corresponding to $N = 60$. Asymptotic height of the potential is the same for all models of this class, see (\ref{energylevel}).}
\label{tmodelfig}
\end{figure}

Functions ${\tanh}^{2n}(\varphi/\sqrt6)$ are symmetric with respect to $\varphi \to - \varphi$.\footnote{A similar model, exponentially rapidly approaching a constant positive value of its potential in the limit $\phi \to \pm \infty$, was first discussed in the context of supergravity in  \cite{Goncharov:1983mw}.} To study inflationary regime in this model at $\varphi \gg 1$, it is convenient to represent them as follows:
\begin{equation}\label{eframe111a}
V(\varphi) = \lambda \left({1- e^{-\sqrt{2/3}\, \varphi}\over 1+ e^{-\sqrt{2/3}\, \varphi}}\right)^{2n}= \lambda \left(1- 4n\, e^{-\sqrt{2/3}\, \varphi} + O\left(n^{2}\, e^{-2\sqrt{2/3}\, \varphi}\right)\right).
\end{equation}

The slow-roll equation for the field $\varphi$ at $\varphi \gg 1$ in terms of the large e-folding number $N$ is
\be
{d\varphi\over dN} = {V'\over V} = 4n \sqrt{2\over 3}\ e^{-\sqrt{2/3}\, \varphi} \ .
\ee\label{eqN}
For large $N$, this leads to a relation
\be
e^{-\sqrt{2/3}\, \varphi(N)} = {3\over 8n\, N} \ .
\ee
Therefore for a given $N$, one has
\be
 {V'\over V} = 4n  \sqrt{2\over 3}\ e^{-\sqrt{2/3}\, \varphi(N)}=   { 3\over 2N} \sqrt{2\over 3} \ ,
\ee
and therefore
\be
\epsilon = {1\over 2} \left({V'\over V}\right)^{2} = {1\over 2} \Big (4n \sqrt{2\over 3}\ e^{-\sqrt{2/3}\, \varphi(N)}\Big )^2={1\over 2} \Big ({ 3\over 2N}  \sqrt{2\over 3} \Big )^2\ .
\ee
This result, in the leading order in $1/N$, is valid for any $n$. The same is true for the slow roll parameter $\eta$, and, consequently, for the parameters $n_{s}$ and $r$:  for the set of T-Models described in this section,  
\be
1 -n_{s} =2/N\, , \qquad r = 12/N^{2}
\ee
 in the leading approximation in $1/N$. 

One could expect that these results may become increasingly unreliable for large $n$, but in fact the expansion parameter is $1/N$ for each model, so the difference of the predictions for   $1-n_{s} =2/N$ and $r = 12/N^{2}$ in the slow-roll approximation is indeed $O(1/N^{2})$. We checked this statement by explicitly comparing models ${\tanh}^{2}(\varphi/\sqrt6)$ and ${\tanh}^{20}(\varphi/\sqrt6)$, and we found, numerically, that in the slow-roll approximation in both cases $n_{s} \approx 0.967$ and $r \approx 0.0032$ for $N \approx 60$.\footnote{We write $N \approx 60$ because the end of inflation can be defined in several different ways, which differ from each other by $\Delta N = O(1)$.}

It will be important for the future discussion that for each of these models, the points were the perturbations are produced on the scale corresponding to $60$ e-foldings corresponds to the same height ${\Delta V\over V} = {3\over 2 N} \sim 1/40$, where $\Delta V = V(\varphi_{\infty})-V(\varphi_{60})$. The line ${\Delta V\over V} \sim 1/40$ is shown by red dashed line in Fig. \ref{tmodelfig}.

The scalar potential in the T-Model is equal to a positive cosmological constant in an infinitely large region $-\infty < \varphi < +\infty$, everywhere except a finite interval $O(5)$ near $\varphi = 0$. Therefore initial conditions for inflation in this class of models appear to be quite natural, especially in the context of the string landscape scenario. For a discussion of initial conditions for inflation in models of this type, see e.g. \cite{Linde:1987yb}.

\begin{figure}[ht!]
\centering
\includegraphics[scale=0.99]{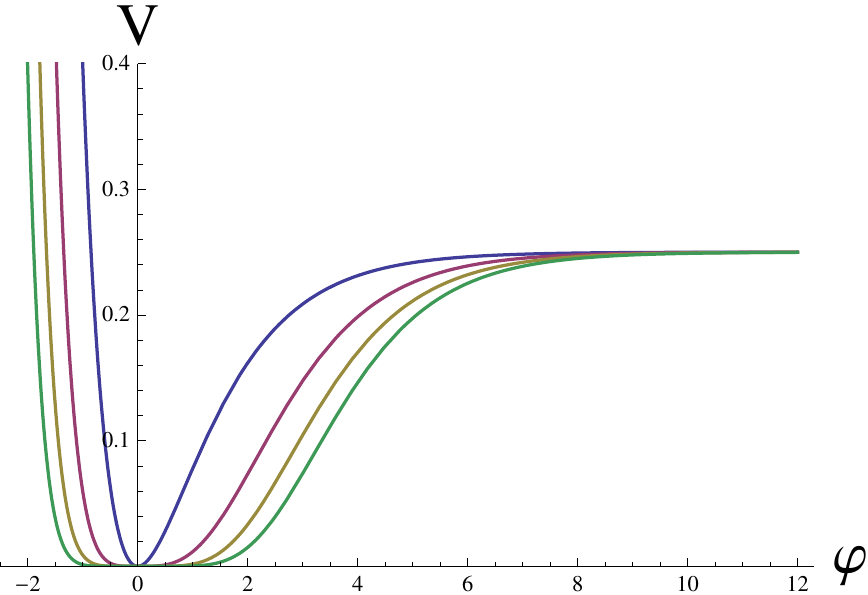}
\caption{Models of conformal inflation based on generalizations of the Starobinsky model, with $F({\phi/\chi}) \sim {\phi^{2n}\over \chi^{2n-2}(\phi+\chi)^{2}}$, $n = 1,2,3,4$.}
\label{stretchstar}
\end{figure}

As we have shown in our previous paper \cite{Kallosh:2013lkr},  Starobinsky model \cite{Starobinsky:1980te} also belongs to the class of chaotic conformal inflation models discussed above, with $F({\phi/\chi}) \sim {\phi^{2}\over (\phi+\chi)^{2}}$, which leads to 
\be
V(\varphi)\sim   \Big [{ \tanh (\varphi/\sqrt6)) \over 1+ \tanh (\varphi/\sqrt6) ) }\Big ]^2 \sim  \left(1- e^{-\sqrt{2/3}\, \varphi}\right)^{2} \ .
\ee
From the perspective of conformal inflation, Starobinsky model with $F({\phi/\chi}) \sim {\phi^{2}\over (\phi+\chi)^{2}}$ it somewhat more complicated and asymmetric  than the basic T-Model (\ref{TModel}) with $F({\phi/\chi}) \sim {\phi^{2}\over \chi^{2}}$. However, all of these models, as well as their generalizations such as $F({\phi/\chi}) \sim {\phi^{2n}\over \chi^{2n-2}(\phi+\chi)^{2}}$, Fig. \ref{stretchstar}, give the same predictions for $n_{s}$ and $r$.

The reasons for this universality, and further generalizations of the models with such properties will be discussed in the next section.

\section{Universality of conformal inflation}

In this section we will describe the roots of the universality of predictions of conformal inflation in a more general way. But first of all, we will consider some instructive examples, which will help to explain the main idea of our approach.

\begin{figure}[ht!]
\centering
\includegraphics[scale=1]{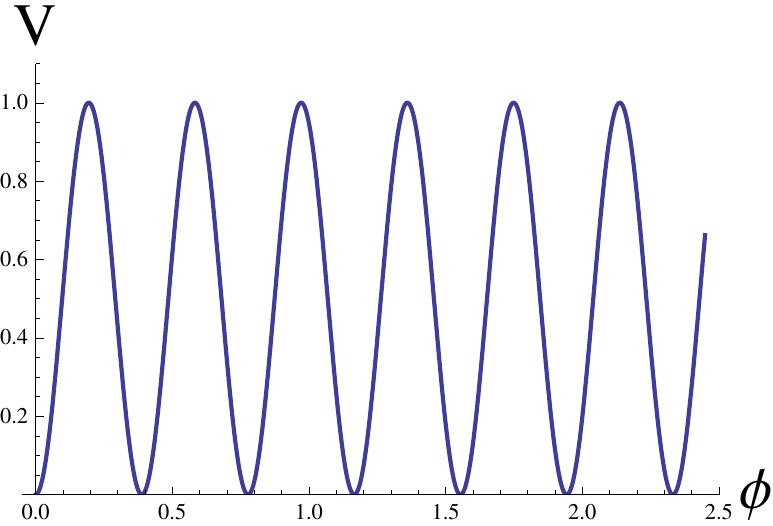}\hskip 0.8 cm \includegraphics[scale=1]{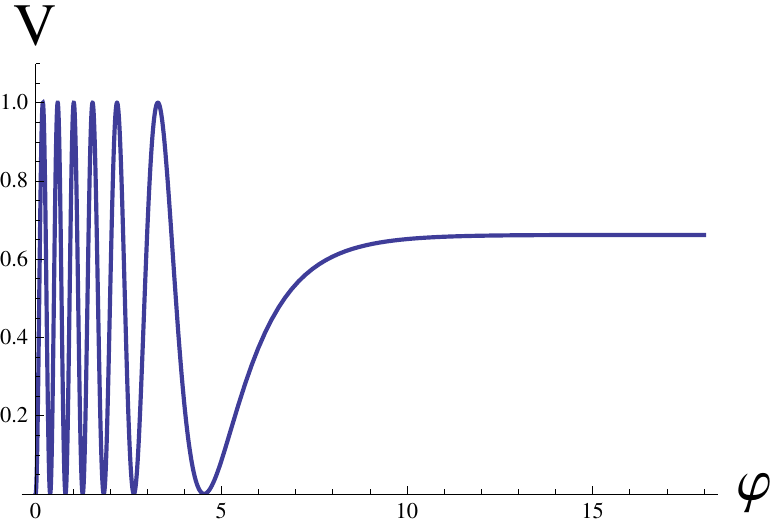}
\caption{Flattening of the sinusoidal potential $V(\phi)$ near the boundary of the moduli space $\phi = \sqrt 6$ by boost  in the moduli space, $V(\phi) \to V(\sqrt 6 \, \tanh {\varphi\over \sqrt{6}})$. Inflationary plateau of the function $V(\phi)$ appears because of the exponential stretching of the last growing part of the sinusoidal function $V(\phi)$.}\label{stretchsin}
\end{figure}

Consider a sinusoidal function $F\left({\phi/\chi}\right) \sim \sin(a+ b \phi)$ and check what will happen to it after the boost $V(\phi) \to V(\sqrt 6 \, \tanh {\varphi\over \sqrt{6}})$. As we see from the Fig. \ref{stretchsin}, the main part of the stretch of the potential occurs very close to the boundary of moduli space, near $\phi = \sqrt 6$. The rising segment of the sinusoidal function bends and forms a plateau, which has an ideal form for the slow-roll inflation. However, if the last segment of the sinusoidal function were falling down, its stretching would produce an  exponentially decreasing curve rapidly approaching  dS space. This possibility does not lead to slow roll inflation ending in a nearly Minkowski space in which we live now. Thus if we consider such functions, we have a 50\% chance that their stretching will produce a nice inflationary potential.

\begin{figure}[ht!]
\centering
\includegraphics[scale=1]{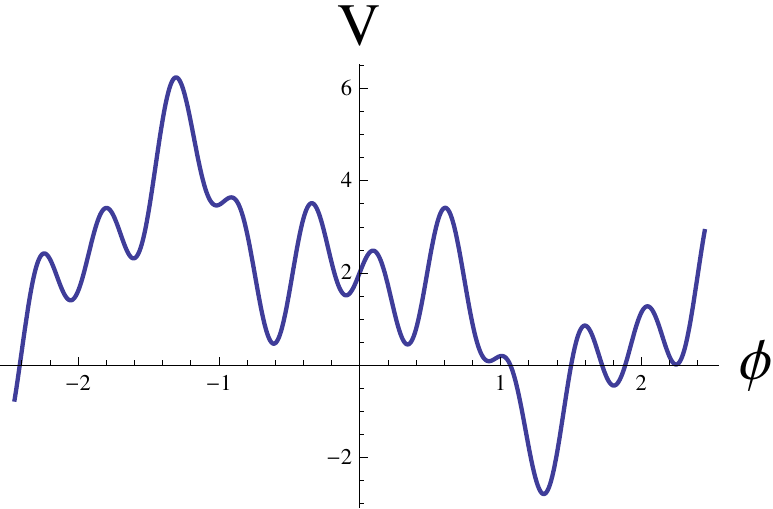}\hskip 0.8 cm \includegraphics[scale=1]{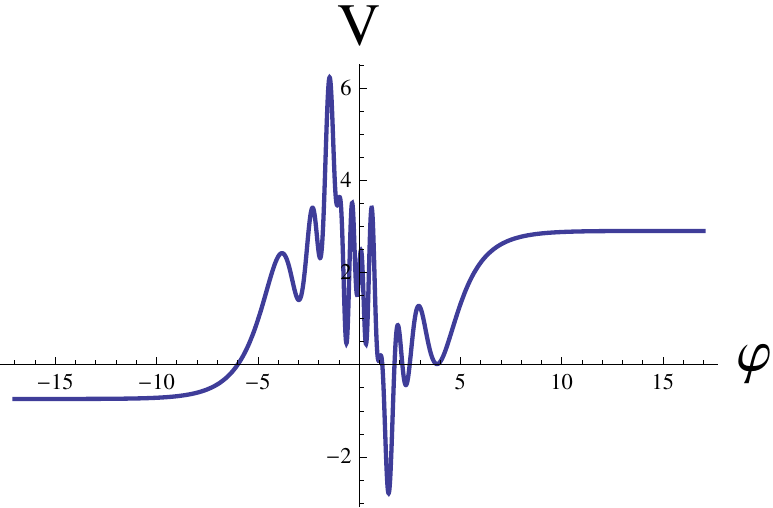}
\caption{Flattening of a generic potential near the boundary of the moduli space by boost  in the moduli space $V(\phi) \to V(\sqrt 6 \, \tanh {\varphi\over \sqrt{6}})$. In essence, what we see is `inflation {\it of} the inflationary landscape' at the boundary of the moduli space, which solves the flatness problem of the inflationary potential required for inflation {\it in} the landscape. }
\label{stretchchaotic}
\end{figure}

Now let us study a more general and complicated potential  on the full interval $-\sqrt 6 < \phi< \sqrt 6$, as shown in Fig. \ref{stretchchaotic}. It shows the same effect as the one discussed above: The part of the landscape at $|\phi| \ll \sqrt 6$ does not experience any stretching. The part with a falling field $\phi$  near the left boundary of the moduli space at $\phi = -\sqrt 6$ becomes $dS$ or (as in Fig. \ref{stretchchaotic}) AdS space. The inflationary plateau appears because of the exponential stretching of the growing branch of $V(\phi)$ near $\phi = \sqrt 6$.

In this scenario, inflationary regime emerges each time when $V(\phi)$ grows at the boundary of the moduli space, which is a rather generic possibility, requiring no exponential fine-tuning.\footnote{As usual, there is an unavoidable fine-tuning of the cosmological constant, which is the value of the potential at the local minimum corresponding to our part of the universe. We presume that this issue can be taken care off by the usual considerations involving inflationary multiverse and string theory landscape, see e.g. \cite{Linde:1984ir,Bousso:2000xa}.}

To specify the conditions required for this scenario to work, as well as its observational consequences, we will represent the potential $V(\phi)$ by expanding it in Taylor series near the boundary of the moduli space, in terms of the deviation $x$ of the field $\phi$ from $\sqrt 6$: \, 
\be \label{x}
x =  1  - {\phi\over \sqrt 6} = \left(1- \tanh {\varphi\over \sqrt 6}\right) \approx 2\  e^{-\sqrt{2/3}\, \varphi}\  ,
\ee
\be\label{expansion}
V(\varphi) = V_{*}\ (1 - \sum  c_{n} x^{n} )= V_{*}\ \left[1 - \sum  c_{n} \left(2\  e^{-\sqrt{2/3}\, \varphi}\right)^{n} \right] \ , 
\ee
where $V_{*}\equiv V(x=0)$, which is equal to the asymptotic value of $V(\varphi)$ at $\varphi\to \infty$. We will try to solve this equation keeping the first term in the expansion,
\be
V(\varphi) = V_{*}\ \left(1 - 2\, c_{1}\  e^{-\sqrt{2/3}\, \varphi} \right) \ ,%= V_{*}\, (1 - \sum  c_{n} (\sqrt 6 - \phi)^{n} )
\ee
assuming that $2\, c_{1}\  e^{-\sqrt{2/3}\, \varphi}  \ll 1$, and then we will estimate the contribution of higher terms.

In this approximation, equation for the field $\varphi$ in the slow roll approximation is the same as in eq.~(\ref{eqN}), with an obvious replacement $4n \to 2\, c_{1}$.
For large $N$, this leads to relation
\be
2\, c_{1} \, e^{-\sqrt{2/3}\, \varphi(N)} = {3\over 2\, N} \ ,
\ee
where $\varphi_{N}$ is the value of the inflaton  field  at the time corresponding to $N$ e-foldings from the end of inflation.

This implies that
\be
\Delta V = V_{*}- V(\varphi_{N}) = {3\over 2\, N} \ , 
\ee
independently of $c_{1}$, see Fig. \ref{tmodelfig}. More importantly, it implies that expansion in powers of $x$ in (\ref{expansion}) is, in fact, expansion in $1/N$. Therefore, unless the slope of the potential $c_{1}$ is anomalously small, or some of the coefficients $c_{n}$ are anomalously large, one can indeed ignore all higher corrections for $N\gg 1$, and all models with these properties (and $c_{1} > 0$) will have identical values of $n_{s}$ and $r$, independently of the detailed structure of the model. This is the main reason for the universality which we found in many models of conformal inflation described above.

Flattening of the potentials that we have found is similar but different from the corresponding effect in the theory $\lambda\phi^{4} -  {\xi\over 2}\phi^{2} R$ with $\xi > 0$ \cite{Salopek:1988qh}, \cite{Sha-1}. However, for $\xi > 0$ this effect is efficient  only for theories $\sim \phi^{4}$, whereas in our case the effect is more universal. The results of our approach can be further extended, for example, to the description of models with arbitrary value of $\xi < 0$, see e.g. a discussion of one of such models in \cite{Linde:2011nh}. The main reason for flattening of the inflationary potential in the models discussed in this paper can be explained as follows. 

In the original conformally invariant theory, the only conformally invariant measure of the amplitude of the scalar field $\phi$ is the ratio $z = {\phi\over \chi}$. After fixing $\chi = \sqrt 6$, which brings us to the Jordan frame, the variable $z = {\phi\over \chi}$ becomes ${\phi\over \sqrt 6}$. One can use the variable $x = 1-z = 1- {\phi\over \sqrt 6}$, which we introduced in the previous section as a measure of the distance from the boundary of the moduli space at $\phi = \sqrt 6$.
 
Gradients of the potential $V$ in terms of the canonically normalized field $\varphi$ are related to gradients in terms of $x$:
\be
V_{\varphi} = {dV\over d\varphi} = {dV\over dx} {dx\over d\varphi}= \sqrt{2\over 3}\, x\, {dV\over dx}
=  {dV\over dx} \times 2\ \sqrt{2\over 3}\  e^{-\sqrt{2/3}\, \varphi} \ , \ee 
see eq. (\ref{x}). 

This result shows that the flatness of the potential $V(\varphi)$ (the smallness of ${dV\over d\varphi}$) can be interpreted as a result of the stretching of the moduli space, which occurs when one goes from variables $z$, $x$ or $\phi$ to the canonically normalized field $\varphi$ in the Einstein frame.
At small $x$ (large $\varphi$), one has
\be
d\varphi   = {1\over 2}\,\sqrt{3\over 2}\, e^{\sqrt{2/3}\, \varphi}\  dx \ .
\ee
Thus, any small step by $dx$ in the original conformal variables near the boundary of the moduli space translates into an exponentially longer step  ${1\over 2}\sqrt{3\over 2} e^{\sqrt{2/3} \varphi}$ in terms of $\varphi$. This effect
 is responsible for the flattening of the inflationary potential in terms of the canonical field $\varphi$. It is reminiscent of two different effects from other areas of physics: Exponential stretching of inhomogeneities and decrease of their gradients during inflation, and redshift of light (increase of its wavelength) emitted by an observer falling towards the black hole horizon.
 
 From these results, one can derive an additional universal parameter, the energy scale of inflation, $V \approx V_{*}$, which takes the same value for all models described above, in the leading order in $1/N$. Planck2013 data imply the following normalization of the amplitude of scalar perturbations of metric \cite{Ade:2013rta}:
\be
{V^{3}\over 12\pi^{2}V^{2}_{\varphi}} \approx 2.2 \times 10^{-9} \ , 
\ee
where $V_{\varphi} = {dV\over d\varphi}$.  In our case, $V(\varphi_{N}) = V_{*} - {3\over 2\, N} \approx V_{*}$, and 
\be
V_{\varphi}= V_{*} \times  2\ c_{1}\, \sqrt{2\over 3}\  e^{-\sqrt{2/3}\, \varphi} =   {3V_{*}\over 2\, N} \ .
\ee
This yields
\be\label{energylevel}
V(\varphi_{N}) =  3.9 \times 10^{-7} N^{-2} \sim 1.1 \times 10^{-10}  
\ee
in Planck units, for all models in this universality class.

\section{Superconformal Universality Class Models}

The superconformal action in general is defined by an embedding \K\, potential ${\cal N}$ and superpotential ${\cal W}$, in notation of \cite{Kallosh:2000ve}
\be
{1\over \sqrt{-g}}\mathcal{L}_{\rm sc}^{\rm scalar-grav}=-\ft16{\cal N} (X,\bar X)R
-G_{I\bar J}{\cal D}^\mu X^I\,{\cal D}_\mu \bar X^{\bar J}-G^{I\bar J}{\cal W}_I \bar{{\cal W}}_{\bar J} \, , \qquad I, \bar I = 0,1,2.
 \label{sc}
\end{equation}
Here $X^0$ is a conformon, $X^1\equiv \Phi $ is an inflaton and $X^3\equiv S$ is a Goldstino, see for example \cite{Kallosh:2013pby} where this setting for our superconformal models is explained in details.
The bosonic universality class models described above has an underlying class of superconformal models:
the embedding \K\, potential for these models has an $SU(1,1)$ symmetry between the complex conformon $X^0$ and the complex inflaton $X^1$ superfields.
\be
\mathcal{N}(X,\bar X)= -|X^0|^2 + |X^1|^2 + |S|^2  - 3 \zeta\,  {(S\bar S)^2\over |X^0|^2-|X^1|^2 }\,  \,.
\label{calNminimal2}
 \ee
We take the superpotential   which preserves a subgroup of $SU(1,1)$, namely an $SO(1,1)$, which makes  a boost between the holomorphic parts of $X^0$ and $X^1$.
 \be
{\cal W} = S \Big ((X^0)^2- (X^1)^2\Big ) f (X^1/X^0)  \ .
 \label{sup}\ee
Function  $f (X^1/X^0)$ is invariant under local conformal-$\mathbb{R}$-symmetry, but if it is not a constant, it breaks the $SO(1,1)$.

\subsection{ Conformal-$\mathbb{R}$-symmetry gauge $X^0= \bar X^0= \sqrt 3$}
In the gauge where we fix the local conformal as well as a local $U(1)$ symmetry by taking $X^0= \bar X^0= \sqrt 3$ and with $X^1\equiv \Phi$ we recover, following  \cite{Ferrara:2010yw,Ferrara:2010in}, a supergravity version of the superconformal model with 
\be
K= -3 \ln \Big [ 1- {|\Phi |^2 + |S^2|\over 3} + \zeta \, {(S\bar S)^2\over 3-|\Phi|^2 } \Big],
\label{K}\ee
and 
 \be
 W = S \Big (3- (\Phi )^2\Big ) f (\Phi/\sqrt 3) \ .
 \ee
An advantage of this gauge is that we may use a well tested over the years  Mathematica code \cite{Kallosh:2004rs}, developed further in \cite{Kallosh:2010ug}, 
and study moduli stabilization. In this case the inflaton is a real part of $\Phi$.
In particular at $S=0$ our condition that $1- {|\Phi |^2 |\over 3}>0$ is a condition that we are inside a `\K\, cone'. The kinetic term for the scalar $\Phi$ is 
\be
\Big({1\over 1-|\Phi |^2 / 3}\Big )^2 \partial_\mu \Phi \partial^\mu \bar \Phi \ .
\ee 
The positivity of the kinetic term for scalars requires that $|\Phi |^2<3$. The boundary of the moduli space here is a `\K\, cone'
\be
1- {|\Phi ^2 |\over 3}=0 \ .
\ee
We find that the condition for stability of goldstino at $S=0$ for all values of the inflaton  is provided by $\zeta>1/6$ in \rf{K}. Inflation is stable at $\rm Im \Phi=0$ independently of the value of $\zeta$.  

\begin{figure}[ht!]
\centering
\vskip 0.2cm \includegraphics[scale=0.5]{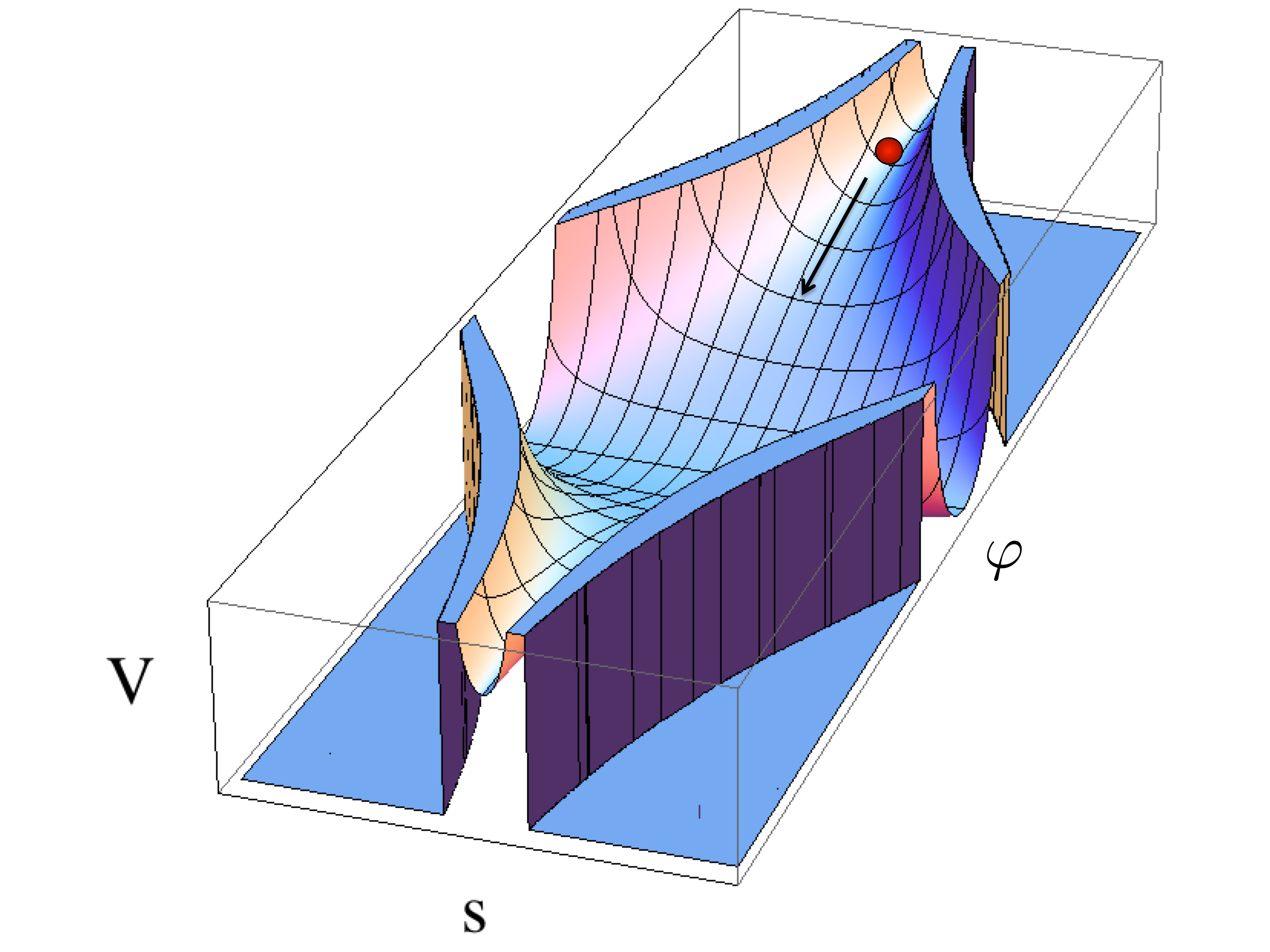}
\caption{Supergravity version of the T-Model. The $S$-direction is steep, the system quickly reaches the minimum at $S=0$ and evolves in the inflaton direction $\varphi$ at stabilized $S$.}
\label{potential}
\end{figure}

As an example of a theory with all moduli stabilized, we show the potential of the fields $\varphi$ and $S$ in the supergravity generalization of the T-Model with $\zeta = 1$ in Fig. \ref{potential}.

\subsection{   $SO(1,1)$ invariant conformal-$\mathbb{R}$-symmetry rapidity gauge $(X^0)^2 - (X^1)^2 = 3$}

Now we use the fact which we learned from the previous subsection, using Mathematica:  inflation takes place at
\be
S=0\, ,  \qquad X^1= \bar X^1= \varphi/\sqrt 2 \ .
\ee
In this gauge we may resolve the $SO(1.1)$ constraint so that 
\be
X^0= \sqrt 3 \cosh   \varphi/ \sqrt 6 \ , \qquad X^1= \sqrt 3 \sinh   \varphi/ \sqrt 6 \ .
\ee
The superconformal action \rf{sc}
with \rf{calNminimal2} and \rf{sup} entries becomes at 
$
S=0 $,  $X^1= \bar X^1= \varphi/\sqrt 2
$
\be
{1\over \sqrt{-g}}\mathcal{L}_{\rm sc}^{\rm scalar-grav}=\ft12  
R
-{1\over 2} (\partial^\mu \varphi)^2 -  |f (\tanh\varphi)|^2 \, .
 \label{scGrav11}
\end{equation}
This shows the origin of the bosonic universality class models in case that $F= f\bar f $.

\section{Emergence of (super)conformal critical phenomena in cosmology}

Numerous studies of supersymmetric attractors in general relativity so far focused on black hole type geometries, where stabilization of moduli near the black holes horizon  was described as a critical phenomena, starting with  \cite{Ferrara:1995ih}  where it was shown that supersymmetric attractors have certain universality classes.

 The superconformal critical phenomena in cosmology  were already introduced in \cite{Kallosh:2013pby}.
 It was observed there that the model ${\lambda\over 4} \phi^4 - {\xi\over 2}\phi^2 R$, when embedded into a corresponding superconformal theory, provides an interpretation of the non-minimal coupling parameter $\xi$ as a deviation from the critical point $\xi=0$. In the superconformal model the relevant parameter $\Delta$ represents a deviation of the embedding \K\, manifold from the flat one. This parameter has one critical point $\Delta_{\rm cr}=0$ corresponding to the $SU(1,1)$ symmetry between the inflaton $X^1$ multiplet and the conformon $X^0$ multiplet. It has another double critical point $\Delta_{\rm cr}=\pm 1/6$ corresponding to an enhanced symmetry between the inflaton and the conformon. At this critical point the parameter of the non-minimal coupling to gravity in the bosonic models becomes equal to the difference of the parameter $\Delta$ from the critical point $\Delta_{\rm cr}=\pm 1/6$, i. e. $\xi= \Delta -\Delta_{\rm cr}$.  This allows to interpret $\xi\geq 2\cdot 10^{-3}$ as a deviation of $\Delta$ from the superconformal critical point $\Delta_{\rm cr}= \pm 1/6$ to fit the Planck data \cite{Ade:2013rta}.
 Note that without the conformal embedding there is no meaning for the parameter $\xi=0$ as a critical point, a point of enhanced symmetry. In presence of the conformon, which is required for the conformal symmetry of the action, the critical points and the corresponding critical phenomena in cosmology can be clearly formulated and studied.
 
 In this paper, the critical phenomena are described with regard to a critical value $\Delta_{\rm cr}=0$, when the  embedding \K\, manifold remains flat and preserves an $SU(1,1)$ symmetry. This symmetry is broken to the $SO(1,1)$ symmetry by the superpotential. The critical point in this case is de Sitter space.
 The deviation from the critical point is introduced via an arbitrary zero conformal weight function $f(z)$  of the homogeneous coordinate $z=X^1/X^0$ in the superpotential. Under Weyl symmetry and $U(1)$ $\mathbb{R}$-symmetry $z$ is invariant, therefore  an arbitrary function $f(z)$
is the only way to describe the deformation of the model near the critical point with $SO(1,1)$ symmetry preserving the local superconformal symmetry of the action. As expected from the general theory of critical phenomena, we find a universality class for such models. What was not known, a priory, is that the attractor point for this universality class models is in fact the one given in eq.   \rf{attractor}.

\section{Conclusions}

We described the emergence of the critical phenomena which arise in the framework of the  superconformal approach to cosmology \cite{Kallosh:2000ve}. This approach helps to classify and generalize a certain class of models favored by Planck. In this paper as well as in \cite{Kallosh:2013pby,Kallosh:2013lkr}
we have shown that the data may be associated with attractor points in the $n_s$-$r$ plane where there a universality class of models with the same (favored by Planck) values of $n_s$ and $r$ (\ref{attractor}) and inflationary energy density (\ref{energylevel}). 

In the context of this approach we identified the effect of exponential flattening of the potentials which appear in a large class of conformal and superconformal models. In these models, the original inflationary potential should be specified in terms of  homogeneous coordinates $z$ that are invariant with respect to (super)conformal transformations. For example, for the conformal theory of the conformon $\chi$ and inflaton $\phi$, this variable is $z = \phi/\chi$. After the gauge fixing and converting into the Einstein frame, the finite interval of variation of $z$ is stretched into an infinite range of the canonical field $\varphi$, which is similar to rapidity in the special theory of relativity. This stretching is similar to inflation which leads to exponential suppression of gradients of inhomogeneities in inflationary universe. In our context, this stretching leads to exponential flattening of the potentials $V(z)$ of the original conformal or superconformal theory, expressed in terms of the canonical field $\varphi$ describing these potentials in the Einstein frame.  This effect facilitates inflation in the theories with a broad class of potentials $V(z)$ and leads to universality of their observational predictions.

It is likely that other interesting developments in inflationary cosmology along the lines developed in \cite{Kallosh:2013pby,Kallosh:2013lkr} and in this paper will follow, and new attractor points in the evolution of the universe will be discovered.

\subsection*{Acknowledgments}

Our work was influenced by discussions with G. `t Hooft about the opportunities to understand physics via spontaneous breaking / alternative  gauge-fixing of conformal symmetry.
We are  grateful  to R. Bond, G. Horowitz, V. Mukhanov,  E. Silverstein, M. Srednicki, L. Susskind,  and participants of the Primordial Cosmology workshop in KITP   for  stimulating discussions. 
This work  is supported by the SITP and by the 
NSF Grant No. 0756174, the work of RK is also supported by the Templeton Foundation Grant ``Frontiers of Quantum Gravity''.

 %\appendix

\section{Appendix: `Special relativity' in the moduli space:  rapidity versus velocity}

The mechanism of flattening the inflationary potentials described in our paper is related to approach to a vicinity of a critical point of  a system at the boundary of the moduli space. It is  geometric and universal for a large class of models.
The physical reason for this phenomena is as deep as the fact that the speed of light $c=1$ cannot be exceeded.
At small velocities the addition rule is
\be
v=v_1+ v_2 \ .
\ee 
Meanwhile, at large velocities the rule is such that 
\be
v= {v_1+v_2\over 1+v_1v_2} \ ,
\ee
so that it never exceeds speed of light. Meanwhile, if we describe the motion using rapidity, instead, the addition rule is
\be
\tanh \theta= {\tanh \theta_1+\tanh \theta_2\over 1+\tanh \theta_1 \tanh \theta_2}= \tanh (\theta_1+\theta_2) \ .
\ee
This means that rapidities have a simple addition rule, 
\be
\theta= \theta_1+\theta_2 \ .
\ee
meanwhile it is  guaranteed that $\tanh$ of any argument never exceeds 1 and speed of light remains the maximal possible velocity.

At small velocities 
\be
\tanh\theta \approx \theta\, , \qquad  (\tanh\theta)^{2n} \approx \theta^{2n} \ .
\ee
However, at large $\theta$ the situation changes dramatically:
\be
(\tanh\theta)^{2n}|_{\theta\rightarrow \infty} \Rightarrow 1\, , \qquad \theta^{2n}|_{\theta\rightarrow \infty}\Rightarrow \infty  \ .
\ee
This is what we find in our potentials, up to the obvious replacement $\theta \to  \varphi/\sqrt 6$, see for example T-Model in Fig. 1. The potential at small $\varphi$ may be more or less steep, it does not matter, they all approach an attractor point at large $\varphi$, independently of their
initial steepness: they comply with `special relativity' in the moduli space.  The major property of the attractor system is that in the process of evolution the system develops to a common final point, independently of initial conditions.

\end{document}